\documentclass{article}%

\usepackage{amsmath,amssymb}
\usepackage{url}
\usepackage{xurl} %
\usepackage{graphics}
\usepackage{algorithm}
\usepackage{algpseudocode}
\usepackage{xcolor}
\usepackage{graphicx}
\usepackage{float}
\usepackage{placeins}
\usepackage{booktabs}
\usepackage{multirow}
\usepackage[title]{appendix}

\title{Curvature-Guided Sparsification for Graphs with Community Structure}

\author{%
Avin Kolahdooz\thanks{Corresponding author.}\\
\small Department of Mathematics and Statistics, University of New Mexico\\
\small \texttt{akolahdooz@unm.edu}
\and
Jos\'e Jairo Garc\'ia\\
\small Department of Mathematics and Statistics, University of New Mexico\\
\small \texttt{josejgarcia@unm.edu}
\and
Sarah Percival\\
\small Department of Mathematics and Statistics, University of New Mexico\\
\small \texttt{spercival@unm.edu}
}

\begin{document}

\maketitle

\begin{abstract}
{Graph sparsification in large-scale networks often requires balancing structural accuracy and computational efficiency. 
Curvature-based methods, particularly Ollivier-Ricci curvature (ORC), have shown strong performance in identifying inter-community bridges, but their computational cost limits scalability. In this work, we propose a two-stage curvature-guided pipeline for graph sparsification with the goal of preserving community structure. We first use the Lower Ricci Curvature, a combinatorial lower bound of the ORC, to perform efficient edge filtering, substantially reducing graph size before applying the computationally intensive optimal-transport-based ORC curvature and Ricci flow.

Results from empirical studies on synthetic and real-world datasets demonstrate that the combined LRC-ORC pipeline can achieve community-recovery accuracy comparable to ORC-only sparsification in terms of Adjusted Rand Index (ARI) and Normalized Mutual Information (NMI), while reducing runtime on most evaluated benchmarks. Our findings suggest that integrating combinatorial and geometric curvature notions provides a computationally efficient and structurally principled framework for graph sparsification in complex networks.}

\end{abstract}

\\

\section{Introduction}

Many real-world structures, such as social \cite{snapfacebook}, biological \cite{Costanzo2010}, or purchase networks \cite{snapamazon}, can be modeled as graphs: nodes represent users, genes, or purchases, and edges model interactions between them. As scientific and computational advances have allowed for the development and study of increasingly large graphs, graph sparsification has become an important tool for distilling large graphs into a more manageable representation. In contrast to graph coarsening, in which both nodes and edges are collapsed, \emph{graph sparsification} refers to the process of removing edges from a graph while retaining the node set, often with the aim of preserving some graph property such as cut weight \cite{Benczur1996} or the spectrum of the graph Laplacian \cite{Spielman2004}. In this work, we focus on graphs that possess a node labeling scheme indicating membership to one or more classes, or communities, and explore graph sparsification methods that reveal community structure.

Graph community structure is closely related to graph topology: intra-community nodes are more likely to be linked by an edge than inter-community nodes. Researchers have observed that graph-based curvatures, derived from the classical notion of curvature on manifolds, capture this community structure well \cite{Tian2025,Ni2019,Sia2019}. On manifolds, Ricci curvature measures the convergence behavior of geodesic paths: in negatively curved space, neighboring geodesics diverge, and in positive curved space, they converge. In the discrete setting, inter-community bridges have negative curvature and edges between highly-connected intra-community nodes have positive curvature. Several generalizations of manifold curvature to discrete spaces have been proposed, including Forman-Ricci curvature (FRC) \cite{Forman2003} and Ollivier-Ricci curvature (ORC) \cite{Ollivier2009ricci}. These notions provide different approaches of characterizing local and global graph geometry. ORC captures geometric information through optimal transport between neighboring node distributions, while FRC determines curvature based on on local combinatorial structure.

The intimate relationship between curvature and community structure suggests that a curvature-based sparsification method which removes highly negatively-curved edges will retain graph communities well. We choose to focus on Ollivier-Ricci curvature since the ORC of a graph converges to the Ricci curvature of the underlying manifold in certain settings \cite{vanderHoorn2021}. 
Although ORC has been shown to perform well in community detection tasks \cite{Tian2025}, it is computationally expensive because optimal-transport calculations must be performed across graph edges, with a computational complexity of $O(md^{5/2})$, where $d$ is the maximum node degree and $m$ is the number of edges \cite{DasGupta2023computingcurvature}. 
To address this computational challenge, Park and Li \cite{Park2025lrc} introduced the lower Ricci curvature (LRC), which provides a fast combinatorial lower bound to ORC; the LRC of an edge can be computed using local combinatorial information in approximately $O(md)$ time. However, because LRC is only an approximation of ORC, LRC-guided graph sparsification does not preserve communities as well as ORC-guided graph sparsification, as demonstrated in Appendix \ref{app:realworld_sensitivity}.

To balance complexity and accuracy, we propose a two-stage sparsification framework which first prunes edges based on the computationally inexpensive LRC, then computes the ORC with on the reduced graph and prunes the graph further based on these refined values. Following the work of \cite{KARAMPOUR2025110302}, we additionally perform a Ricci flow step that introduces edge weights based on the discrete Ricci flow introduced in \cite{Ni2019}, which we use to refine ORC values before the final pruning step.
We validate the proposed framework on synthetic and real-world benchmark networks and analyze its performance in both timing and accuracy tests under different graph structures. 
The remainder of this paper is organized as follows: Section \ref{sec:background} introduces the necessary background and notation. Section \ref{sec:datasets} describes the datasets. Section \ref{sec:methodology} presents the proposed methodology. Section \ref{sec:results} reports the experimental results. We conclude with Section \ref{sec:conclusion}.

\section{Background and Notation}
\label{sec:background}

\subsection{Graph notation and basic operators}

We consider an undirected graph \(G=(V,E,w)\) with node set \(V\), edge set \(E\subseteq V\times V\), and nonnegative edge weights \(w:E\to \mathbb{R}_{+}\). 
For an edge \((i,j)\in E\), we write \(w_{ij}=w_{ji}>0\). We denote the neighborhood of node \(i\) as
\[
N(i)=\{j\in V : (i,j)\in E\},
\]
and its weighted degree as
\[
d_i=\sum_{j\in N(i)} w_{ij}.
\]

\noindent We denote the shortest-path distance between nodes $i$ and $j$ by $d(i,j)$, using hop distance for unweighted graphs and weighted shortest paths after Ricci flow. %

\subsection{Curvature in network geometry}

\begin{figure}[htbp]
\centering
\includegraphics[width=0.7\textwidth]{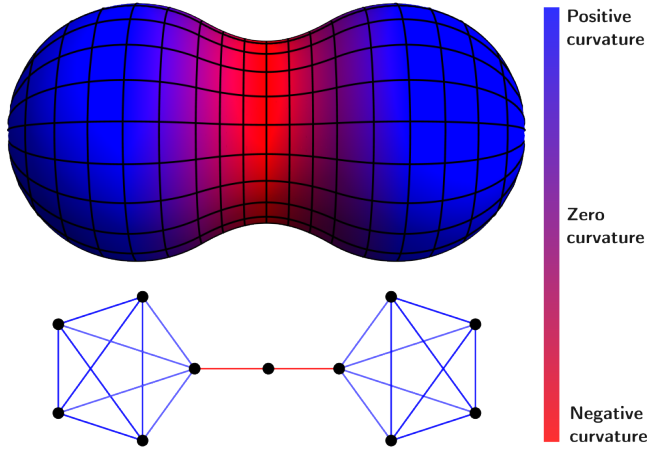}
\caption{Geometric intuition behind curvature-based community detection.
Edges connecting dense communities tend to have negative curvature and act as
bottlenecks, while edges inside dense regions tend to have higher curvature. Image adapted from \cite{Fu2025}.}
\label{fig:curvature_intuition}
\end{figure}

Local graph connectivity can be characterized via graph curvature:
edges within densely connected regions generally exhibit higher curvature,
whereas edges connecting different communities tend to have lower 
curvature.
Figure~\ref{fig:curvature_intuition} illustrates a geometric interpretation
of curvature on both manifolds and graphs.
Several notions of discrete graph curvature have been introduced; we review the ones used in this paper below.

\subsubsection{Ollivier-Ricci curvature (ORC)}

Ollivier-Ricci curvature extends Ricci curvature on manifolds to metric measure spaces via optimal transport \cite{Ollivier2009ricci}. Intuitively, ORC measures the extent to which two random walks from different points converge (positive curvature) or diverge (negative curvature). In the graph setting, we follow \cite{Lin2011} and adopt the  $\alpha$-lazy random walk formulation to define the probability measure at node $i\in V$ as
\[
\mu_i^{(\alpha)} 
= \alpha\,\delta_i 
+ (1-\alpha)\sum_{j\in N(i)} \frac{w_{ij}}{d_i}\,\delta_j,
\]
where \(\delta_i\) denotes the Dirac measure at node \(i\) and \(\alpha\in[0,1]\) is the idleness parameter. This defines a lazy random walk where the walker stays at the current node with probability $\alpha$, and with probability $1-\alpha$, moves to one of the neighboring nodes at random. When \(\alpha=0\), \(\mu_i^{(0)}\) is simply the standard one-step random walk distribution. %
We use the Wasserstein distance to formalize the idea of distances between random walks:
\paragraph{Wasserstein-1 distance.}

Let \(\Pi(\mu_i^{(\alpha)},\mu_j^{(\alpha)})\) denote the set of all couplings of \(\mu_i^{(\alpha)}\) and \(\mu_j^{(\alpha)}\). The Wasserstein-1 distance with respect to the graph metric \(d(\cdot,\cdot)\) is defined as

\[
W_1\!\left(\mu_i^{(\alpha)},\mu_j^{(\alpha)}\right)
=
\min_{\pi\in \Pi(\mu_i^{(\alpha)},\mu_j^{(\alpha)})}
\sum_{u\in V}\sum_{v\in V} d(u,v)\,\pi(u,v).
\]

\paragraph{Definition of ORC.}

The Ollivier-Ricci curvature of an edge \((i,j)\in E\) is defined as

\[
\kappa_{\alpha}(i,j)
=
1-\frac{W_1\!\left(\mu_i^{(\alpha)},\mu_j^{(\alpha)}\right)}{d(i,j)}.
\]
In the case of lazy random walks, the value of $k_\alpha$ ranges from -2 to 1. Negative curvature indicates that neighborhoods are far in transport distance relative to \(d(i,j)\), which often corresponds to inter-community edges and bottlenecks \cite{Sia2019}.

\subsubsection{Discrete Ricci flow on graphs}
\label{sec:Ricciflow}

Ricci flow, originally defined in the continuous setting, deforms a space according to curvature. The authors of \cite{Ni2019} observed that the tendency of Ricci flow to separate space into regions of positive curvature is applicable to community detection in graphs, where intra-community edges often have positive Ollivier-Ricci curvature. This motivated their definition of discrete Ricci flow on graphs, which iteratively updates edge weights based on ORC values. At each step, edge weights are updated using the formula

\[
w_{ij}^{(t+1)}
=
d^{(t)}(i,j)
\left(
1-\kappa_{\alpha}^{(t)}(i,j)
\right),
\]
where $d^{(t)}(i,j)$ is the shortest-path distance under the current edge weights and $\kappa_{\alpha}^{(t)}(i,j)$ is the Ollivier--Ricci curvature at iteration $t$. Edge weights are then normalized and the process is repeated until all neighboring curvatures are within a predetermined tolerance, or after a fixed number of steps. After several iterations, negatively curved edges tend to receive larger weights, while positively curved edges tend to receive smaller weights, aiding curvature-based sparsification \cite{Ni2019}.

\begin{figure}[!h]
\centering
\includegraphics[width=\textwidth]{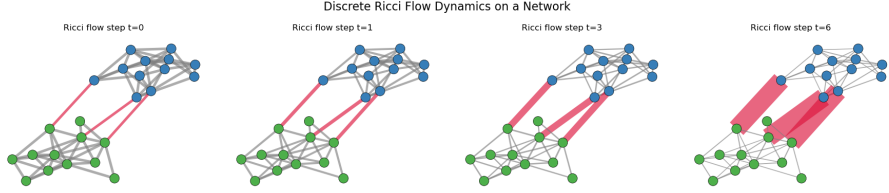}
\caption{Illustration of discrete Ricci flow dynamics. Edge weights
evolve according to curvature: negatively curved edges tend to receive larger weights, while positively curved edges decrease in relative weight, increasing structural contrast.}
\label{fig:ricci_flow}
\end{figure}

\subsubsection{Lower Ricci curvature (LRC)}

Following from the work of \cite{Jost2013}, the authors of \cite{Park2025lrc} introduced the Lower Ricci Curvature (LRC) as computationally inexpensive lower bound on Ollivier--Ricci curvature based on local graph information. For an edge \((i,j)\in E\), we recall that \(N(i)\) denotes the set of neighbors of node \(i\) in the graph. Define

\[
n_i=|N(i)|, \qquad n_j=|N(j)|
\]
and let

\[
n_{ij}=|N(i)\cap N(j)|
\]
denote the number of common neighbors of \((i,j)\). The LRC of \((i,j)\) is defined as

\[
\mathrm{LRC}(i,j)=\frac{2}{n_i}+\frac{2}{n_j}-2+\frac{2n_{ij}}{\max(n_i,n_j)}+\frac{n_{ij}}{\min(n_i,n_j)}.
\]
The value of the LRC ranges from -2 to 2, with lower values indicating weaker local connections.

\subsection{Community detection}

\emph{Community detection} refers to the process of recovering community membership from the underlying graph structure: intra-community nodes are more likely to be linked by an edge than inter-community nodes. Determining class membership from graph topology is a fundamental problem in network science with diverse applications \cite{Flake2000,Girvan2002,Guimer2005}. Many classical community detection algorithms are based on graph clustering; more recently, methods based on diverse areas such as physics and computer science have been introduced \cite{Fortunato2010}. Existing community detection methods include modularity-based algorithms such as the Louvain \cite{Blondel2008louvain} and Leiden \cite{Traag2019leiden} methods; we refer the reader to \cite{Javed2018} for a survey of other techniques. 

Mathematically speaking, community detection partitions a set \(V\), in our case taken to be the nodes of a graph, into \(K\) disjoint communities $C_1,\dots,C_K$ such that intra-community nodes are more likely to be connected by an edge than inter-community nodes. The notion of \emph{modularity} captures this, measuring how many edges fall within communities compared with what would be expected in a random graph with the same degree distribution. In our pipeline, we compute the \emph{weighted} modularity because we use Ricci flow to introduce edge weights (cf.\ Section \ref{sec:Ricciflow}) which strengthen intra-community connections. 

\paragraph{Weighted modularity.}

For a graph with vertex set $V$, we define the total weight as the sum of the edge weights:

\[
m=\frac{1}{2}\sum_{i\in V}\sum_{j\in V} w_{ij}.
\]
Letting $C_i$ be the community label of node $i$, the weighted modularity is 

\[
Q = \frac{1}{2m}\sum_{i\in V}\sum_{j\in V}
\left(w_{ij}-\frac{d_i d_j}{2m}\right)\mathbb{I}(C_i=C_j).
\]
We use the \emph{Louvain method} \cite{Blondel2008louvain}, a modularity-based community detection method, to determine how well sparsified graphs retain their community structure. In short, the Louvain method operates by first assigning each node to its own community, then iteratively grouping nodes, accepting the new group assignments if the (weighted) modularity improves. The algorithm terminates if the modularity stops improving or a predetermined number of iterations has been reached. We fix node evaluation order across datasets using the same set of six seeds.

\subsection{Structural network descriptors}

To characterize the structural separability of our datasets,
we report conductance and empirical within- and between-community edge
densities. These quantities are used only for descriptive analysis and
are not part of the proposed LRC--ORC pipeline.

Let $S \subset V$ be a community and
$\bar{S}=V\setminus S$ its complement.
Let $\operatorname{cut}(S,\bar{S})$ denote
the set of edges connecting $S$ and $\bar{S}$.
The volume of $S$ is defined as
\[
\operatorname{vol}(S)=\sum_{v\in S}\deg(v).
\]
Following \cite{leskovec2009community},
the conductance of $S$ is defined as
\[
\phi(S)=
\frac{|\operatorname{cut}(S,\bar{S})|}
{\min\{\operatorname{vol}(S),
\operatorname{vol}(\bar{S})\}},
\]
Lower conductance indicates stronger
separation between a community and
the rest of the network.

In a graph with $n$ vertices, the maximum number of possible edges is $\binom{n}{2}$. We define the within-community edge density to be the ratio of the number of extant edges within a communities to the total possible number of intra-community edges:

\[
\hat p_{\mathrm{in}}
=
\frac{m_{\mathrm{in}}}{\sum_i \binom{n_i}{2}},
\]
where $n_i$ is the size of ground-truth community $C_i$, and
$m_{\mathrm{in}}$ is the total number of edges between nodes within the same community. Similarly, letting $m_{\mathrm{out}}$ be the number of edges between nodes in different communities, we define between-community edge density as
\[
\hat p_{\mathrm{out}}
=
\frac{m_{\mathrm{out}}}
{\binom{n}{2}-\sum_i \binom{n_i}{2}},
\]
Larger separation
between $\hat p_{\mathrm{in}}$ and $\hat p_{\mathrm{out}}$ indicates that nodes within communities are more likely to be connected to each other than to nodes in other communities. 

\subsection{External validation metrics: ARI and NMI}

Because reference labels are available in each of our datasets (see Section \ref{sec:datasets}), we quantify agreement between a detected partition \(\mathcal{C} = \{C_1, \ldots, C_K\}\) and the ground-truth partition \(\mathcal{G} = \{G_1, \ldots, G_K\}\) using the Adjusted Rand Index (ARI) and Normalized Mutual Information (NMI).

\paragraph{Adjusted Rand Index (ARI).}

The Rand index \cite{Rand1971} measures agreement between two clusterings of data, with a value of 0 if the clustering do not agree on any point, and 1 if they are the same. The adjusted Rand index (ARI) adjusts this measurement to compensate for similarity expected by chance \cite{Hubert1985ari}. Specifically, let \(n_{ab}\) number of elements assigned to community \(C_a\) and \(G_b\),  \(n_{a\cdot}\) the number of elements assigned to $C_a$, and, \(n_{\cdot b}\) the number of elements assigned to $G_b$. Letting $n$ be the total number of vertices in the graph, the ARI is defined as

\[
\mathrm{ARI} =
\frac{
\sum_{a,b}\binom{n_{ab}}{2}
-
\frac{\sum_a\binom{n_{a\cdot}}{2}\sum_b\binom{n_{\cdot b}}{2}}{\binom{n}{2}}
}{
\frac{1}{2}\left[\sum_a\binom{n_{a\cdot}}{2}+\sum_b\binom{n_{\cdot b}}{2}\right]
-
\frac{\sum_a\binom{n_{a\cdot}}{2}\sum_b\binom{n_{\cdot b}}{2}}{\binom{n}{2}}
},
\]

\paragraph{Normalized Mutual Information (NMI).}

The concept of mutual information, introduced in \cite{Shannon1948}, measures how much information can be gained about one random variable by observing another. Formally, let \(U\) and \(V\) be the random variables induced by partitions \(\mathcal{C}\) and \(\mathcal{G}\). Define
\[
p(a,b)=n_{ab}/n, \quad
p(a)=n_{a\cdot}/n, \quad
p(b)=n_{\cdot b}/n.
\]
The mutual information of $U$ and $V$ is defined as
\[
I(U;V)=\sum_{a}\sum_{b} p(a,b)\log\frac{p(a,b)}{p(a)p(b)}.
\]
and the entropies are
\[
H(U)=-\sum_{a} p(a)\log p(a), 
\qquad
H(V)=-\sum_{b} p(b)\log p(b).
\]

The normalized mutual information used in community detection is

\[
\mathrm{NMI}(U,V)=\frac{2I(U;V)}{H(U)+H(V)},
\]

which takes values in \([0,1]\) with larger values indicating stronger agreement.
\section{Datasets}
\label{sec:datasets}

To evaluate the proposed curvature-guided framework, we consider five benchmark datasets spanning both synthetic and real-world networks. These datasets cover a wide range of structural properties, including social, collaboration, and product co-purchase networks, allowing us to investigate how graph structure influences the effectiveness of curvature-guided sparsification.
Since several benchmarks contain overlapping communities, each node is assigned to its smallest ground-truth community for evaluation using ARI and NMI.

\subsection{Synthetic Stochastic Block Model (SBM)}

We generate synthetic benchmark graphs using the Stochastic Block Model (SBM), where nodes are partitioned into predefined communities with dense intra-community connections and sparse inter-community links. By controlling the intra-community probability $p_{in}$ and inter-community probability $p_{out}$, SBM provides a controlled environment for evaluating community detection performance under different structural regimes. Ground-truth community assignments are known by construction.

\subsection{SNAP Facebook Ego-Networks}

The Facebook ego-network dataset from the Stanford Network Analysis Project (SNAP) \cite{snapfacebook,leskovec2012circles} represents friendship relations centered around individual users. Nodes correspond to Facebook users and undirected edges represent friendship links between them. The original graph contains 4,039 nodes and 88,234 edges. Although the original friendship circles allow overlapping community memberships, for evaluation we assign each node a single ground-truth label to obtain a non-overlapping partition compatible with the clustering metrics used in this study. For nodes belonging to multiple circles, we assign the node to its smallest ground-truth community.

\subsection{SNAP Amazon Co-purchase Network}

The Amazon co-purchase network from SNAP \cite{snapamazon,yang2012community} models relationships between products frequently purchased together. Nodes represent products, while edges connect products that are commonly co-purchased by customers. The original graph contains 334,863 nodes and 925,872 edges. Product categories provide overlapping ground-truth communities, making this dataset particularly suitable for evaluating community detection in dense product networks. For evaluation with ARI and NMI, each node is assigned to its smallest ground-truth community to obtain a non-overlapping reference partition.

\subsection{SNAP DBLP Collaboration Network}

The DBLP collaboration network from SNAP \cite{snapdblp,yang2012community} represents scientific collaborations between authors. Nodes correspond to authors and undirected edges indicate co-authorship relationships. The graph contains 317,080 nodes and 1,049,866 edges. Ground-truth communities are defined by publication venues, allowing authors to belong to multiple research communities simultaneously. For evaluation with ARI and NMI, each node is assigned to its smallest ground-truth community to obtain a non-overlapping reference partition.

\subsection{SNAP YouTube Social Network}

The YouTube social network from SNAP \cite{snapyoutube,leskovec2009community} represents friendship relations between users of the YouTube platform. Nodes correspond to users and edges represent friendship links. The complete network contains 1,134,890 nodes and 2,987,624 edges. User-created groups serve as overlapping ground-truth communities, providing a challenging network with substantial community overlap and heterogeneous connectivity. For evaluation with ARI and NMI, each node is assigned to its smallest ground-truth community to obtain a non-overlapping reference partition.

\section{Methodology}
\label{sec:methodology}

\subsection{Computational environment}
All experiments were conducted using \texttt{Python 3.11} \cite{python}.
Graph manipulation and preprocessing were implemented using \texttt{NetworkX} \cite{networkx}.
Lower Ricci curvature computations were implemented using custom routines based on the formulation of Park and Li \cite{Park2025lrc}, using the number of common neighbors and the degrees of adjacent nodes.
Ollivier-Ricci curvature and discrete Ricci flow were computed using the \texttt{GraphRicciCurvature} library \cite{Ni2019python}.
Community detection was carried out primarily using the Louvain algorithm via the \texttt{python-louvain} package \cite{pythonlouvain}. Experiments using the Leiden method were also performed; the complete Leiden results are available in the accompanying GitHub repository and are omitted from the main text for brevity.
Evaluation metrics (ARI and NMI) were computed using \texttt{scikit-learn} \cite{scikit-learn}.
All runtimes were measured as elapsed wall-clock time. For randomized community-detection runs, each method was evaluated using six seeds. Results are reported as mean $\pm$ standard deviation across these runs.
The codebase was tested successfully on (i) a Pop!\_OS 24.04 LTS x86\_64 workstation with 62.49\,GB RAM, an Intel\textsuperscript{\tiny\textregistered} Xeon\textsuperscript{\tiny\textregistered} W-2235 CPU, and an AMD Radeon Pro WX 3200 GPU, and (ii) a MacBook Air running macOS Sonoma 14.8.5 with 8\,GB RAM and an Apple M3 (10-core integrated GPU). In both cases, experiments were run locally using the Conda environment specified in environment.yml in the GitHub repository \cite{KolahdoozKolahdooz2026repo}.

\subsection{Graph representation and preprocessing}

We model all benchmark networks as undirected, initially unweighted
graphs $G=(V,E)$. Self-loops are removed before dataset-specific
benchmark construction.
For the Facebook, DBLP, and YouTube datasets, we first select ground-truth communities of moderate size and conductance, construct the induced subgraph on the union of
their nodes, and retain its giant connected component (GCC), when
necessary, to obtain a connected benchmark.

For the Amazon dataset, we instead restrict the graph to nodes belonging to
eligible ground-truth communities, evaluate the resulting connected
components using size, label coverage, and overlap characteristics,
and select one eligible connected component as the final benchmark.

\subsection{Combined LRC-ORC pipeline}

The proposed method integrates LRC and ORC with Ricci flow (ORC-RF) into a two-stage
pipeline.
LRC is used as a computationally inexpensive preprocessing step, while ORC and Ricci flow are applied only on the reduced graph. The resulting computation is more efficient than running ORC with Ricci flow on the entire graph, yet largely preserves structural information.
The key novelty of the proposed framework is that LRC is not used as an alternative to ORC, but as an efficient graph reduction stage that limits the expensive ORC+Ricci flow computation to structurally informative edges.

The procedure works as follows: we first compute the LRC value of every edge, then remove the edges with the lowest LRC values. The proportion of removed edges, and hence the LRC threshold, depends on the structure of the dataset, which we discuss in more detail in Section \ref{sec:results}. This results in a coarsely sparsified graph that preserves within-community
connectivity while eliminating weak inter-community edges.
Following the discrete Ollivier--Ricci flow framework of Ni et al.~\cite{Ni2018}, 
Ollivier--Ricci curvature is computed on the reduced graph using the Sinkhorn distance as an approximation to the computationally expensive 1-Wasserstein distance on a one-step 
random walk distribution with idleness parameter $\alpha=0.5$.
We perform $T=5$ Ricci-flow iterations for the real-world benchmarks and $T=3$ iterations for the synthetic SBM experiments as computationally tractable settings that limit repeated ORC calculations while allowing curvature-based refinement. 
After LRC pruning, ORC-RF refinement is applied to the largest connected component. The remaining components are retained without further ORC refinement.

Edge weights are normalized after each iteration,
and persistently low-curvature edges are subsequently
pruned. Since Ricci flow also modifies edge weights
used by Louvain, differences in clustering accuracy
may reflect both edge pruning and weight refinement.
Algorithm~1 summarizes the proposed two-stage
curvature pipeline.

\begin{algorithm}[H]
\caption{Two-stage curvature-guided community detection (LRC $\rightarrow$ ORC-RF)}
\label{alg:lrc_orc_pipeline}
\begin{algorithmic}[1]
\Require Graph $G=(V,E)$, prune ratios $p_{\mathrm{LRC}}, p_{\mathrm{ORC-RF}} \in(0,1)$, idleness $\alpha$, Ricci-flow steps $T$
\Ensure Communities $\mathcal{C}$

\State \textbf{LRC stage:} compute $\mathrm{LRC}(e)$ for all $e\in E$
\State $G_1 \gets \textsc{PruneBottom}(G,p_{\mathrm{LRC}})$

\State \textbf{ORC-RF stage:} compute $\kappa_\alpha(e)$ for all $e\in E(G_1)$
\For{$t=1$ to $T$}
    \State update edge weights by Ricci flow using $\kappa_\alpha$
    \State normalize edge weights
    \State recompute $\kappa_\alpha$ on updated weights
\EndFor
\State $G_2 \gets \textsc{PruneBottom}(G_1,\kappa_\alpha,p_{\mathrm{ORC-RF}})$

\State $\mathcal{C} \gets \textsc{Louvain}(G_2)$
\State \Return $\mathcal{C}$
\end{algorithmic}
\end{algorithm}
\FloatBarrier

\section{Results}
\label{sec:results}

The following experiments compare the proposed LRC$\rightarrow$ORC-RF pipeline with LRC-based and ORC-RF-based component baselines in terms of clustering accuracy, runtime, and graph reduction. Since the role of LRC-based preprocessing has been described in Section~4, the discussion below focuses on dataset-specific performance and structural behavior.

To examine the interaction between the two pruning stages, we performed
a joint sensitivity analysis on the SBM benchmark. As shown in
Figure~\ref{fig:joint_pruning_sensitivity}, moderate pruning generally
preserves clustering quality better than aggressive sparsification. The LRC-only and ORC-RF-only configurations are used as component baselines to isolate the contribution of each stage of the proposed pipeline. We additionally performed experiments using the ORC without Ricci flow, detailed in Appendix \ref{app:ricci_flow_ablation}. Removing Ricci flow substantially reduced runtime, but its effect on accuracy was dataset-dependent. Ricci flow improved community recovery on DBLP and Amazon, but reduced accuracy on Facebook and YouTube.

After pruning, community detection is performed using the Louvain algorithm. In some cases, a small number of nodes were isolated after pruning, which were removed before community detection.
Ground-truth community assignments are available for all benchmark datasets and are used for external validation.
We evaluate performance using the Adjusted Rand Index (ARI) and
Normalized Mutual Information (NMI). Reported runtimes includes curvature computation (with Ricci flow where applicable), and pruning. For each benchmark, the reported conductance corresponds to the average conductance of the ground-truth communities.

\subsection{Synthetic SBM Benchmark}

We generated synthetic networks with 360 nodes divided
into four equally sized communities, with approximately
4,000 edges per realization. The within- and
between-community connection probabilities were
$p_{\mathrm{in}}=0.15$ and $p_{\mathrm{out}}=0.022$,
respectively. We additionally introduced 10 dense
inter-community bridge blocks of size $8\times8$
to create challenging community structures.

Table~\ref{tab:synthetic_results} compares the three
methods across 10 independent graph realizations.
LRC-only was the fastest but had the lowest
accuracy (ARI $=0.766$, NMI $=0.761$).
ORC-RF-only achieved the highest accuracy
(ARI $=0.836$, NMI $=0.818$).
The combined LRC$\rightarrow$ORC-RF approach achieved
similar accuracy (ARI $=0.834$, NMI $=0.816$)
while reducing runtime by approximately $8\%$
relative to ORC-RF-only.

\begin{table}[H]
\centering
\caption{Performance comparison of curvature-based methods
on the synthetic SBM benchmark. Results are reported as
mean $\pm$ standard deviation across 10 runs.}
\label{tab:synthetic_results}
\begin{tabular}{lccc}
\hline
Method & ARI & NMI & Time (s) \\
\hline
LRC-only
& $0.766 \pm 0.080$
& $0.761 \pm 0.058$
& $0.07 \pm 0.01$ \\

ORC-RF-only
& $0.836 \pm 0.094$
& $0.818 \pm 0.068$
& $9.01 \pm 0.40$ \\

LRC$\rightarrow$ORC-RF
& $0.834 \pm 0.087$
& $0.816 \pm 0.066$
& $8.36 \pm 0.34$ \\
\hline
\end{tabular}
\end{table}

\subsubsection{Pruning Sensitivity}

\begin{figure}[H]
\centering

\begin{minipage}{0.48\textwidth}
    \centering
    \includegraphics[width=\linewidth]{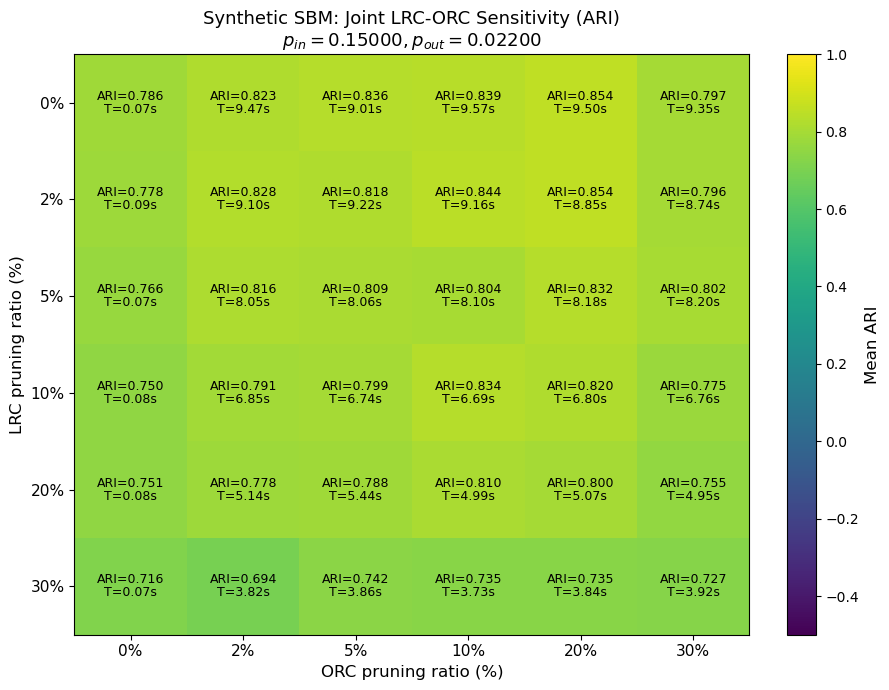}
    \small (a) Adjusted Rand Index (ARI)
\end{minipage}
\hfill
\begin{minipage}{0.48\textwidth}
    \centering
    \includegraphics[width=\linewidth]{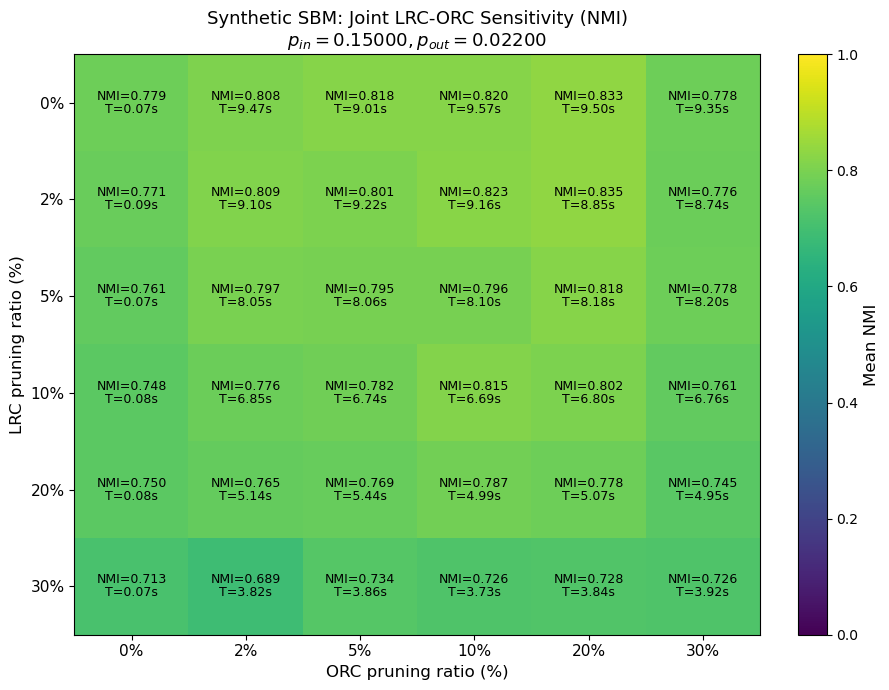}
    \small (b) Normalized Mutual Information (NMI)
\end{minipage}

\caption{Joint sensitivity of (a) ARI and (b) NMI
to LRC and ORC-RF pruning on the synthetic SBM
benchmark.
Results are averaged over 10 independent graph
realizations.}

\label{fig:joint_pruning_sensitivity}
\end{figure}

We evaluated LRC and ORC-RF pruning levels of
$0$, $2$, $5$, $10$, $20$, and $30\%$, using
$\alpha=0.5$ and three Ricci flow iterations.
Figure~\ref{fig:joint_pruning_sensitivity}
shows the mean ARI and NMI across 10 graph realizations.
Both metrics generally decreased under aggressive LRC
pruning, while moderate ORC-RF pruning improved recovery.

\begin{table}[H]
\centering
\caption{Sensitivity of the combined LRC--ORC-RF pipeline
to selected pruning configurations and ORC idleness
parameters on the synthetic SBM benchmark.}
\label{tab:synthetic_sensitivity}
\begin{tabular}{cccccc}
\hline
$p_{\mathrm{LRC}}$ (\%) &
$p_{\mathrm{ORC}}$ (\%) &
$\alpha$ &
ARI & NMI & Time (s) \\
\hline
0  & 10 & 0.50 & 0.838 & 0.816 & 9.060 \\
10 & 30 & 0.25 & 0.788 & 0.765 & 5.104 \\
20 & 25 & 0.75 & 0.751 & 0.735 & 5.370 \\
25 & 30 & 0.75 & 0.710 & 0.709 & 4.682 \\
30 & 20 & 0.50 & 0.718 & 0.707 & 3.619 \\
\hline
\end{tabular}
\end{table}

\noindent
Figure~\ref{fig:joint_pruning_sensitivity} shows that
moderate pruning generally preserves high community-recovery
accuracy, while aggressive pruning tends to reduce both
ARI and NMI.

\subsection{DBLP Collaboration Network}
\label{sec:dblp_results}

We first evaluate the proposed framework on the DBLP
collaboration network, a real-world benchmark containing
4,508 nodes, 12,863 edges, and 25 ground-truth communities.
This benchmark provides an opportunity to examine whether
an inexpensive local-curvature preprocessing step can
reduce the computational cost of Ollivier--Ricci curvature
(ORC) refinement while maintaining community-recovery
accuracy. Table~\ref{tab:dblp_results} compares LRC-only,
ORC-only, and the proposed LRC$\rightarrow$ORC
pipeline.

\begin{table}[H]
\centering
\small
\begin{tabular}{lcccc}
\hline
Method & ARI & NMI & Runtime (s) & Graph Size \\
\hline
LRC-only
& $0.329 \pm 0.012$
& $0.748 \pm 0.003$
& $\mathbf{0.10 \pm 0.00}$
& 4508V / 12219E \\

ORC-RF-only
& $\mathbf{0.493 \pm 0.014}$
& $\mathbf{0.767 \pm 0.007}$
& $27.68 \pm 0.00$
& 4500V / 12219E \\

LRC$\rightarrow$ORC-RF
& $0.395 \pm 0.006$
& $0.757 \pm 0.002$
& $22.76 \pm 0.00$
& 4507V / 12266E \\
\hline
\end{tabular}
\caption{Community detection results on DBLP over
six random seeds (mean $\pm$ standard deviation).
LRC-only and ORC-RF-only use 5\% edge pruning.
The combined pipeline uses 2.5\% LRC pruning
followed by 2.5\% ORC pruning.
Graph size denotes the numbers of vertices (V)
and edges (E) remaining after sparsification.
Runtime follows the preprocessing-time measurement
reported by the experimental pipeline.}
\label{tab:dblp_results}
\end{table}

On DBLP, ORC-only achieves the highest accuracy
(ARI = $0.493$, NMI = $0.767$), while the combined
LRC$\rightarrow$ORC pipeline reduces runtime by
approximately 17.8\% (from 27.68 to 22.76 seconds),
at the cost of lower ARI ($0.395$) and NMI ($0.757$).
LRC-only is considerably faster (0.10 seconds)
but achieves the lowest accuracy. These results suggest that early LRC pruning may
remove structurally informative edges, reducing
community-recovery accuracy. Thus, while the
combined approach offers computational savings,
ORC-only provides better accuracy on DBLP,
highlighting the trade-off between efficiency
and preserving important network structure.

\subsection{Amazon Co-purchasing Network}
\label{sec:amazon_results}

We next consider the Amazon co-purchasing network,
using a selected connected benchmark containing
298 nodes, 1,159 edges, and 9 ground-truth labels
derived from overlapping product communities.
The evaluation uses the benchmark's final
ground-truth labeling.

Unlike DBLP, the Amazon benchmark allows us to
investigate a setting in which the combined
LRC$\rightarrow$ORC-RF approach achieves accuracy
comparable to the ORC-only method while
reducing its computational cost.

\begin{table}[htbp]
\centering
\small
\begin{tabular}{lcccc}
\hline
Method & ARI & NMI & Runtime (s) & Graph Size \\
\hline
LRC-only
& $0.440 \pm 0.003$
& $0.728 \pm 0.004$
& $\mathbf{0.03 \pm 0.00}$
& 298V / 1101E \\

ORC-only
& $\mathbf{0.519 \pm 0.000}$
& $\mathbf{0.751 \pm 0.000}$
& $1.71 \pm 0.00$
& 298V / 1101E \\

LRC$\rightarrow$ORC
& $\mathbf{0.519 \pm 0.000}$
& $\mathbf{0.751 \pm 0.000}$
& $1.34 \pm 0.00$
& 298V / 1105E \\
\hline
\end{tabular}
\caption{Community detection results on Amazon over
six random seeds (mean $\pm$ standard deviation).
LRC-only and ORC-RF-only use 5\% edge pruning,
while the combined pipeline applies 2.5\%
LRC pruning followed by 2.5\% ORC pruning.
Graph size reports the final numbers of
vertices (V) and edges (E).
Runtime follows the preprocessing-time measurement
reported by the experimental pipeline.}
\label{tab:amazon_results}
\end{table}

As shown in Table~\ref{tab:amazon_results},
LRC$\rightarrow$ORC achieves the same mean ARI
($0.519$) and NMI ($0.751$) as ORC-only,
while reducing runtime by approximately 21.6\%
(from 1.71 to 1.34 seconds). Both methods retain
a similar number of edges (1,105 and 1,101,
respectively), suggesting that the combined approach
reduces computational cost without sacrificing
community-recovery accuracy.

Although LRC-only is considerably faster
(0.03 seconds), it achieves lower ARI ($0.440$)
and NMI ($0.728$), highlighting the benefit of
ORC-based refinement on this network.
Unlike DBLP, where preliminary LRC pruning
reduces accuracy, the Amazon results show that
the combined pipeline can preserve ORC-level
performance at a lower computational cost.

\subsection{Facebook Network}
\label{sec:facebook_results}

We further evaluate the framework on a benchmark
derived from Facebook friendship circles.
The largest connected component of the selected
benchmark contains 1,771 nodes, 42,220 edges,
and 39 ground-truth communities.

The Facebook benchmark exhibits heterogeneous
community structure. The mean conductance
of its final ground-truth communities is
approximately $0.4944$, indicating that many
communities are not sharply separated from
the surrounding network. This setting allows us to examine whether
the more computationally expensive ORC
refinement provides an advantage over local
curvature-based sparsification.

\begin{table}[htbp]
\centering
\small
\begin{tabular}{lcccc}
\hline
Method & ARI & NMI & Runtime (s) & Graph Size \\
\hline
LRC-only
& $\mathbf{0.649 \pm 0.006}$
& $\mathbf{0.769 \pm 0.003}$
& $\mathbf{0.44 \pm 0.00}$
& 1771V / 40109E \\

ORC-only
& $0.585 \pm 0.028$
& $0.738 \pm 0.010$
& $82.36 \pm 0.00$
& 1771V / 40109E \\

LRC$\rightarrow$ORC
& $0.586 \pm 0.002$
& $0.733 \pm 0.001$
& $77.31 \pm 0.00$
& 1770V / 40137E \\
\hline
\end{tabular}
\caption{Community detection results on Facebook
over six random seeds (mean $\pm$ standard deviation).
LRC-only and ORC-only prune 5\% of edges.
The combined method applies 2.5\% LRC pruning
followed by 2.5\% ORC pruning.
Graph size indicates the numbers of vertices (V)
and edges (E) after sparsification.
Runtime follows the preprocessing-time measurement
reported by the experimental pipeline.}
\label{tab:facebook_results}
\end{table}

As shown in Table~\ref{tab:facebook_results},
LRC-only achieves the highest ARI ($0.649$)
and NMI ($0.769$), while requiring only
0.44 seconds. ORC-only and
LRC$\rightarrow$ORC achieve lower ARI values
of $0.585$ and $0.586$, respectively.
Although the combined pipeline reduces runtime
by approximately 6.1\% compared with ORC-RF-only
(from 82.36 to 77.31 seconds), it does not
improve clustering accuracy.

These results suggest that the local structural
information captured by LRC is sufficient for
effective community recovery on Facebook,
and additional ORC refinement provides no
further benefit under the evaluated settings.
Unlike DBLP and Amazon, where ORC-based methods
outperform LRC-only, Facebook illustrates that
more computationally expensive curvature
refinement does not necessarily lead to
better community detection.

\subsection{YouTube Network}
\label{sec:youtube_results}

Finally, we evaluate the framework on a
YouTube social-network benchmark containing
1,760 nodes, 5,600 edges, and 23 ground-truth
communities. This benchmark provides another opportunity
to examine whether the additional computational
cost of ORC refinement leads to improved
community-recovery accuracy.

\begin{table}[htbp]
\centering
\small
\begin{tabular}{lcccc}
\hline
Method & ARI & NMI & Runtime (s) & Graph Size \\
\hline
LRC-only
& $\mathbf{0.676 \pm 0.041}$
& $\mathbf{0.800 \pm 0.009}$
& $\mathbf{0.05 \pm 0.00}$
& 1760V / 5320E \\

ORC-only
& $0.619 \pm 0.016$
& $0.742 \pm 0.014$
& $7.21 \pm 0.00$
& 1756V / 5320E \\

LRC$\rightarrow$ORC
& $0.622 \pm 0.017$
& $0.753 \pm 0.006$
& $7.36 \pm 0.00$
& 1759V / 5323E \\
\hline
\end{tabular}
\caption{Community detection results on YouTube
over six random seeds (mean $\pm$ standard deviation).
LRC-only and ORC-only use 5\% edge pruning.
The combined pipeline uses 2.5\% LRC pruning
followed by 2.5\% ORC pruning.
Graph size denotes the numbers of vertices (V)
and edges (E) after sparsification.
Runtime follows the preprocessing-time measurement
reported by the experimental pipeline.}
\label{tab:youtube_results}
\end{table}

As shown in Table~\ref{tab:youtube_results},
LRC-only achieves the highest ARI ($0.676$)
and NMI ($0.800$) among the three curvature-based
methods, with a runtime of only 0.05 seconds.
ORC-only and LRC$\rightarrow$ORC achieve
lower ARI values of $0.619$ and $0.622$,
respectively, with corresponding NMI values
of $0.742$ and $0.753$. The combined method
is also slightly slower than ORC-only
(7.36 versus 7.21 seconds).

\begin{figure}[htbp]
\centering
\includegraphics[width=\textwidth]{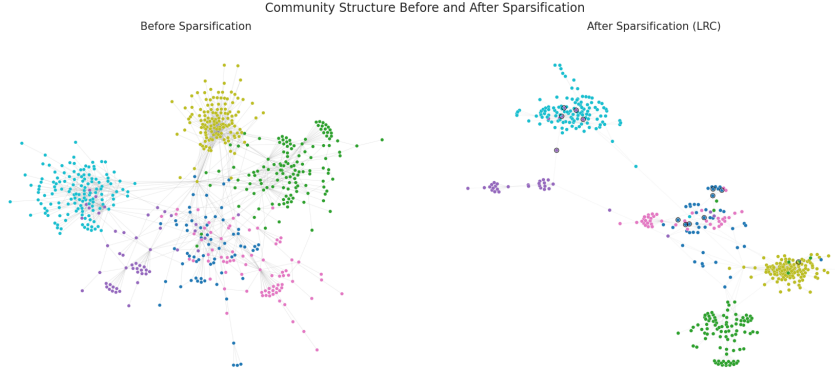}
\caption{Illustration of community structure
before and after LRC-based sparsification on the YouTube
benchmark. The visualization highlights
changes in graph connectivity following
edge removal; quantitative community-recovery
performance is reported separately in
Table~\ref{tab:youtube_results}.}
\label{fig:graph_example}
\end{figure}

These results suggest that local structural
information is sufficient for effective community
recovery on YouTube, while additional ORC
refinement offers no improvement under the
evaluated settings. However, the unpruned Leiden
baseline achieves a higher ARI ($0.705$),
indicating that LRC-only is the best of the
three curvature-based methods, rather than
the best overall approach.
Figure~\ref{fig:graph_example} illustrates
the changes in community structure before
and after sparsification.

\subsection{Structural Regimes of
Curvature-Guided Community Detection}
\label{sec:structural_regimes}

Our experiments show that no single curvature-guided
sparsification method consistently outperforms the others
across all benchmarks. To better understand these differences,
we compare the results using two measures of community
separation: mean conductance and the empirical ratio of
within- to between-community edge probabilities,
$\hat p_{\mathrm{in}}/\hat p_{\mathrm{out}}$.
Lower conductance generally indicates stronger community
separation, while a higher probability ratio indicates
denser connections within communities relative to those
between communities.

\begin{table}[htbp]
\centering
\small
\begin{tabular}{lccc}
\hline
Dataset
& Mean Conductance
& $\hat p_{\mathrm{in}}/\hat p_{\mathrm{out}}$
& Best Accuracy Method \\
\hline
Synthetic & 0.410 & 4.35  & ORC-only \\
DBLP      & 0.040 & 552.5 & ORC-only \\
Amazon    & 0.449 & 25.8  & ORC-only / Combo \\
Facebook  & 0.494 & 44.9  & LRC-only \\
YouTube   & 0.243 & 25.5  & LRC-only \\
\hline
\end{tabular}
\caption{Structural characteristics and performance of
curvature-guided methods across benchmarks. The best
method is determined by mean ARI among LRC-only,
ORC-RF-only, and LRC$\rightarrow$ORC-RF.}
\label{tab:structural_regimes}
\end{table}

As shown in Table~\ref{tab:structural_regimes},
neither conductance nor
$\hat p_{\mathrm{in}}/\hat p_{\mathrm{out}}$
alone explains which method performs best.
DBLP has strongly separated communities but favors
ORC-only, whereas Facebook, with weaker community
separation, achieves higher accuracy using LRC-only.
Similarly, Amazon and YouTube have nearly identical
within-to-between community probability ratios
(25.8 and 25.5), yet favor different approaches.
These differences suggest that community separation
alone does not capture the structural information
needed to determine which edges can be safely removed.

The value of ORC refinement also varies across networks.
On DBLP, ORC-only provides the highest accuracy,
while on Amazon, the combined pipeline maintains
comparable accuracy at a lower computational cost.
In contrast, LRC-only performs best among the
curvature-based methods on Facebook and YouTube,
where additional ORC-RF refinement offers no
improvement under the evaluated settings.

The experiments without Ricci flow, reported in
Appendix~\ref{app:ricci_flow_ablation}, further support
this observation. Removing Ricci flow reduces
runtime, but its effect on accuracy differs across
datasets, indicating that iterative refinement
is beneficial in some settings but unnecessary
in others.

Overall, these findings highlight a trade-off
between computational efficiency and preserving
structurally informative edges. While the proposed
LRC$\rightarrow$ORC pipeline can reduce the cost
of ORC-based processing without sacrificing accuracy
on some networks, early pruning may reduce community
recovery on others. The results therefore suggest
that effective curvature-guided sparsification
depends on finer network structure beyond global
community-separation measures, rather than
following a universally optimal strategy.

\section{Conclusion and Future Work}
\label{sec:conclusion}

This work introduced a two-stage curvature-guided sparsification framework that uses LRC-based pruning before ORC-RF refinement. The proposed pipeline can reduce the computational cost of Ollivier-Ricci curvature with Ricci flow on most evaluated benchmarks, while maintaining competitive community detection performance
on selected networks. The results further show that the effectiveness of curvature-guided sparsification depends on network structure. Overall, the study demonstrates that combinatorial and geometric curvature can be used as complementary components of an efficient and structure-aware community detection framework.

One limitation is that the initial LRC-based sparsification step may result in the removal of all edges between two communities, limiting the effectiveness of our method in applications that require graph connectivity. Future work will investigate adaptive pruning strategies that account for graph connectivity and reduce the risk of removing too many cross-community edges. We additionally would like to explore the sparsification regimes delineated in several of the heatmaps in Appendix \ref{app:realworld_sensitivity}. Coarse measurements of graph structure alone do not account for this behavior; it remains an open question as to which factors determine the presence of these accuracy drop-offs.

\begin{appendices}

\section{The Effect of Ricci Flow}
\label{app:ricci_flow_ablation}

To assess the contribution of Ricci flow, we compare
the combined LRC--ORC pipeline with and without
iterative Ricci-flow refinement. Both configurations
use 2.5\% LRC pruning followed by 2.5\% ORC pruning,
allowing us to examine the effect of Ricci flow
on community-recovery accuracy and computational cost.

\begin{table}[htbp]
\centering
\small
\setlength{\tabcolsep}{4pt}
\begin{tabular}{llccccc}
\toprule
Dataset & Configuration
& LRC (\%) & ORC (\%)
& ARI & NMI & Time (s) \\
\midrule

\multirow{2}{*}{DBLP}
& With Flow
& 2.5 & 2.5
& $0.395 \pm 0.006$
& $0.757 \pm 0.002$
& 22.76 \\
& No Flow
& 2.5 & 2.5
& $0.302 \pm 0.011$
& $0.738 \pm 0.003$
& 3.95 \\

\midrule

\multirow{2}{*}{Amazon}
& With Flow
& 2.5 & 2.5
& $0.519 \pm 0.000$
& $0.751 \pm 0.000$
& 1.34 \\
& No Flow
& 2.5 & 2.5
& $0.436 \pm 0.000$
& $0.724 \pm 0.000$
& 0.292 \\

\midrule

\multirow{2}{*}{Facebook}
& With Flow
& 2.5 & 2.5
& $0.586 \pm 0.002$
& $0.733 \pm 0.001$
& 77.31 \\
& No Flow
& 2.5 & 2.5
& $0.645 \pm 0.003$
& $0.777 \pm 0.001$
& 13.31 \\

\midrule

\multirow{2}{*}{YouTube}
& With Flow
& 2.5 & 2.5
& $0.622 \pm 0.017$
& $0.753 \pm 0.006$
& 7.36 \\
& No Flow
& 2.5 & 2.5
& $0.696 \pm 0.021$
& $0.803 \pm 0.012$
& 1.61 \\

\bottomrule
\end{tabular}
\caption{Comparison of the combined LRC--ORC
pipeline with and without Ricci flow across
four real-world benchmarks. ARI and NMI are
reported as mean $\pm$ standard deviation
over six random seeds where available.}
\label{tab:ricci_flow_ablation}
\end{table}

Removing Ricci flow substantially reduces runtime
across all four benchmarks, but its effect on
community-recovery accuracy varies by dataset.
Ricci flow improves ARI and NMI on DBLP and Amazon,
whereas the combined method without flow achieves
higher scores on Facebook and YouTube.
These results suggest that iterative Ricci-flow
refinement is beneficial for some network structures
but is not universally necessary. The findings
further support the structure-dependent behavior
discussed in Section~\ref{sec:structural_regimes}.

\section{Additional Sensitivity Heatmaps}
\label{app:realworld_sensitivity}

In this section, we provide heatmaps showing timing and accuracy results across different pruning parameters for each dataset. The Adjusted Rand Index (ARI) uses a scale from -0.5 to 1. The Normalized Mutual Information (NMI) uses a scale from 0 to 1.

\begin{figure}[H]
\centering

\begin{minipage}[t]{0.48\textwidth}
    \centering
    \includegraphics[width=\linewidth]{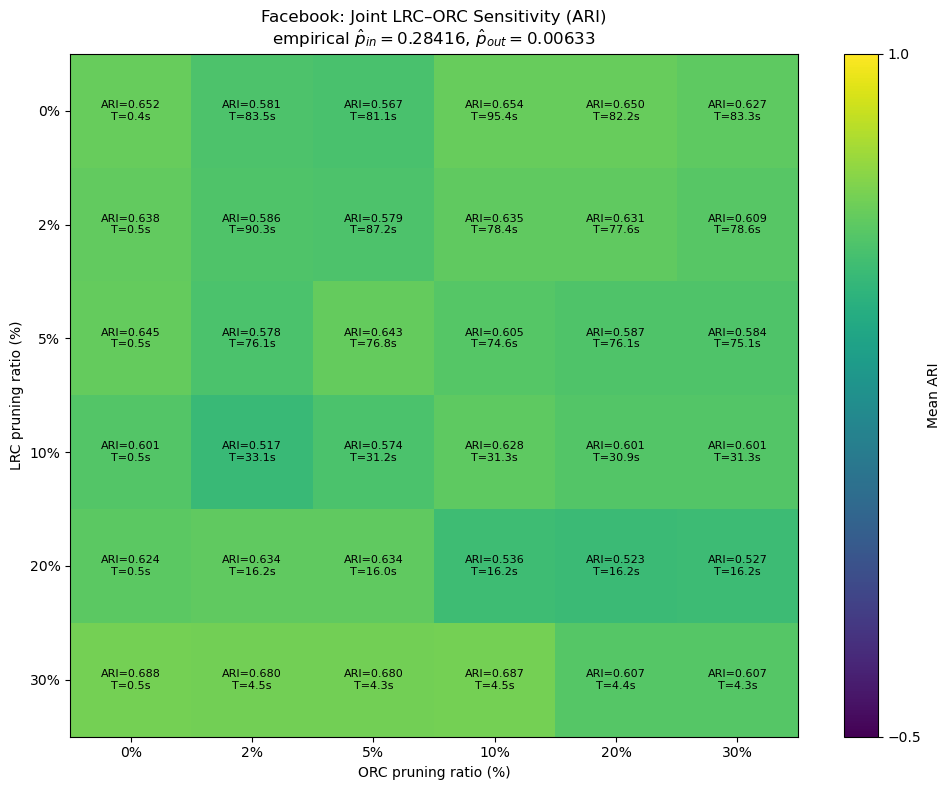 }

    \textbf{(a) ARI}
\end{minipage}
\hfill
\begin{minipage}[t]{0.48\textwidth}
    \centering
    \includegraphics[width=\linewidth]{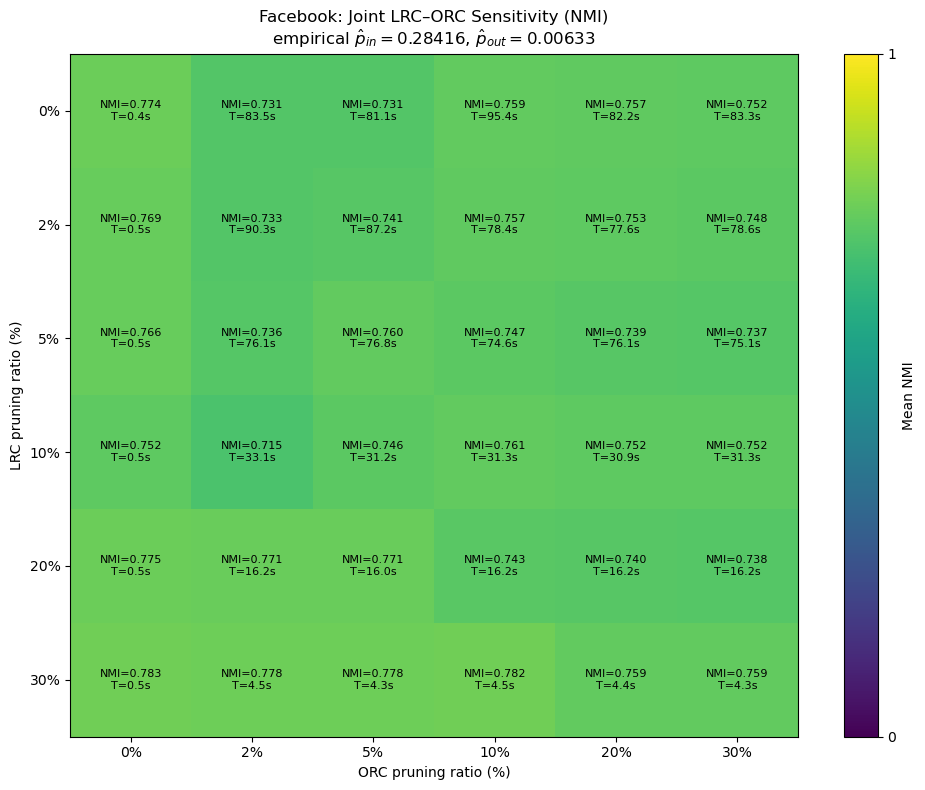}

    \textbf{(b) NMI}
\end{minipage}

\caption{Sensitivity of Facebook community-detection performance to
LRC and ORC-RF pruning parameters.}
\label{fig:appendix_facebook_heatmaps}
\end{figure}

\begin{figure}[H]
\centering

\begin{minipage}[t]{0.48\textwidth}
    \centering
    \includegraphics[width=\linewidth]{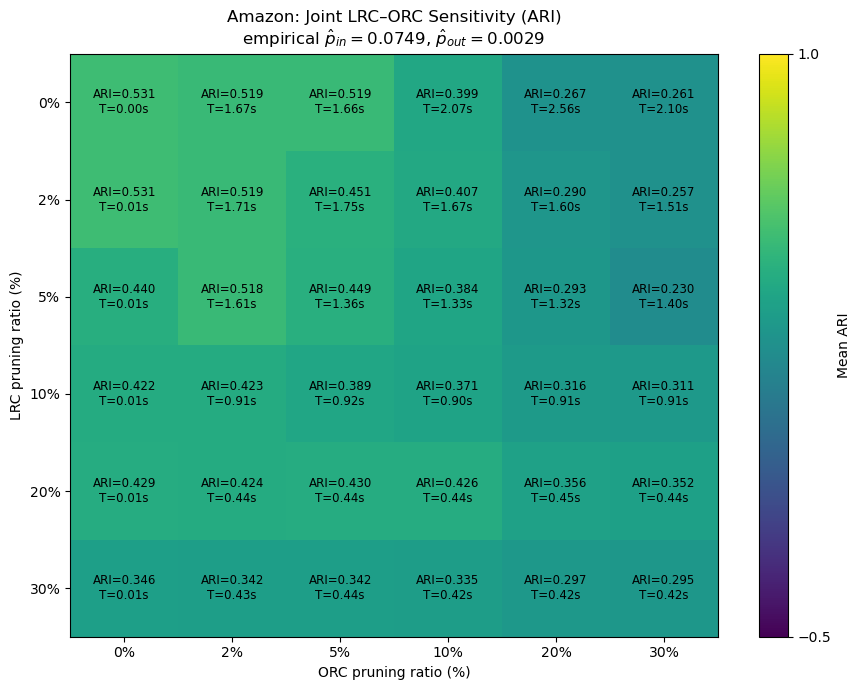}

    \textbf{(a) ARI}
\end{minipage}
\hfill
\begin{minipage}[t]{0.48\textwidth}
    \centering
    \includegraphics[width=\linewidth]{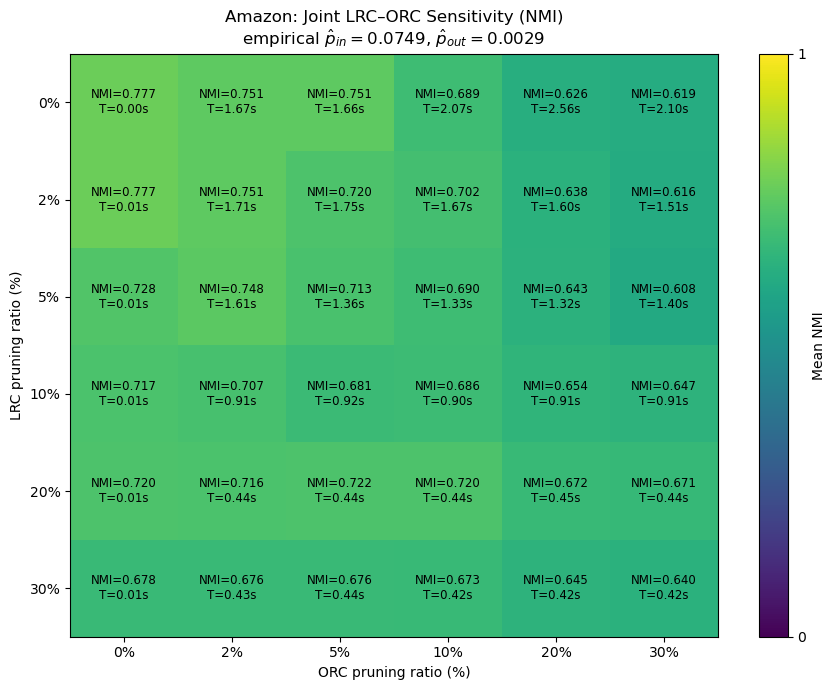}

    \textbf{(b) NMI}
\end{minipage}

\caption{Sensitivity of Amazon community-detection performance to
LRC and ORC-RF pruning parameters.}
\label{fig:appendix_amazon_heatmaps}
\end{figure}

\begin{figure}[H]
\centering

\begin{minipage}[t]{0.48\textwidth}
    \centering
    \includegraphics[width=\linewidth]{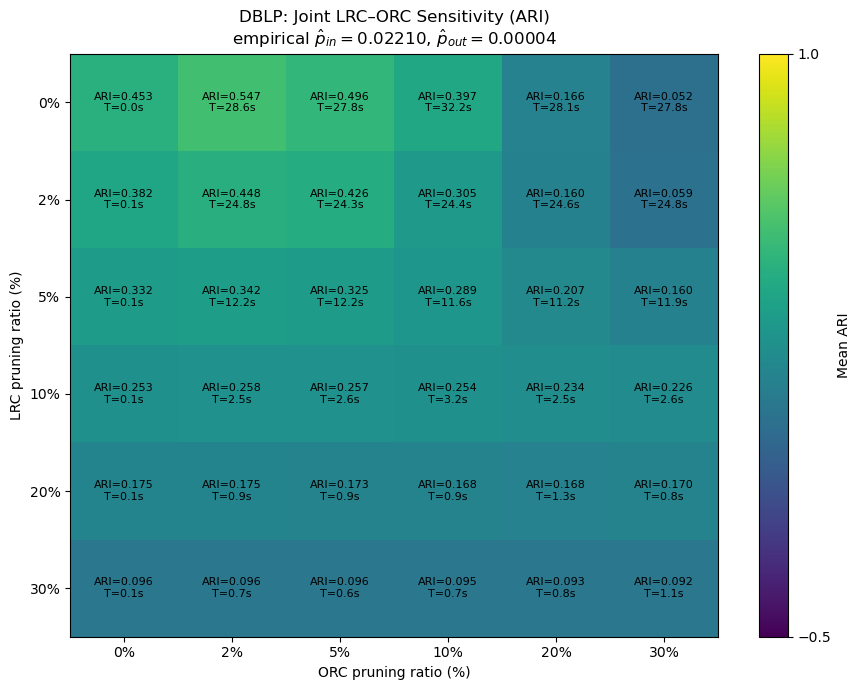}

    \textbf{(a) ARI}
\end{minipage}
\hfill
\begin{minipage}[t]{0.48\textwidth}
    \centering
    \includegraphics[width=\linewidth]{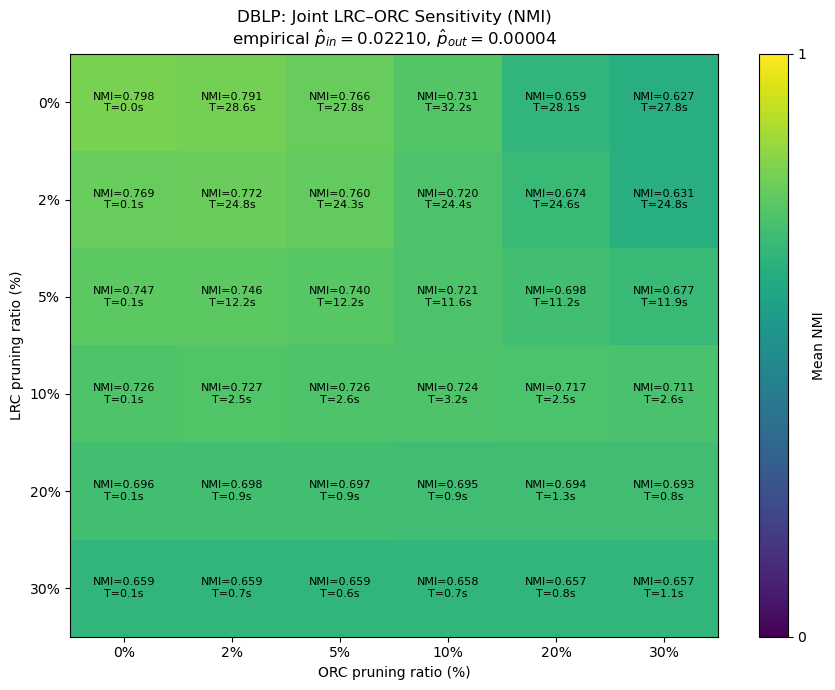}

    \textbf{(b) NMI}
\end{minipage}

\caption{Sensitivity of DBLP community-detection performance to
LRC and ORC-RF pruning parameters.}
\label{fig:appendix_dblp_heatmaps}
\end{figure}

\begin{figure}[H]
\centering

\begin{minipage}[t]{0.48\textwidth}
    \centering
    \includegraphics[width=\linewidth]{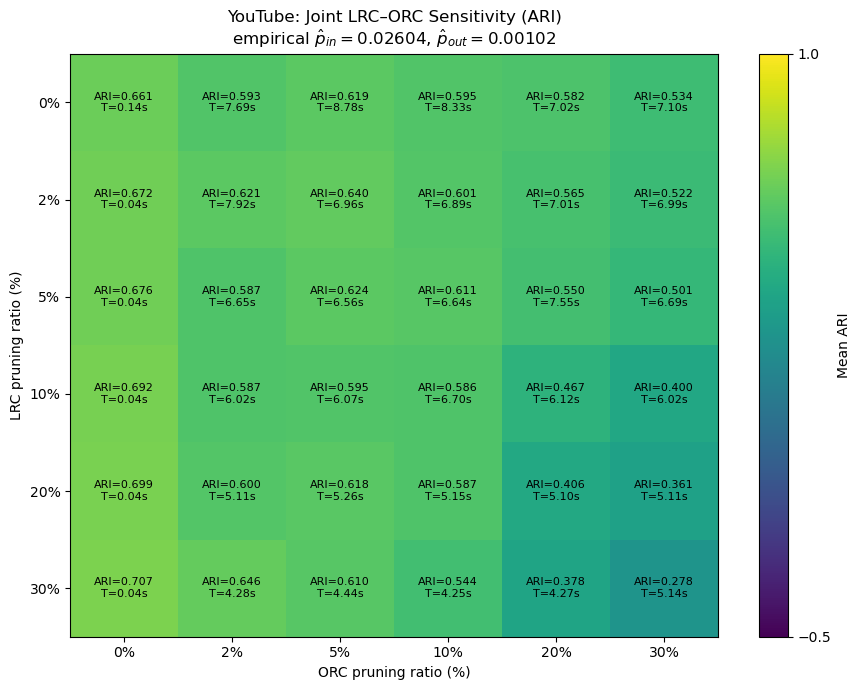}

    \textbf{(a) ARI}
\end{minipage}
\hfill
\begin{minipage}[t]{0.48\textwidth}
    \centering
    \includegraphics[width=\linewidth]{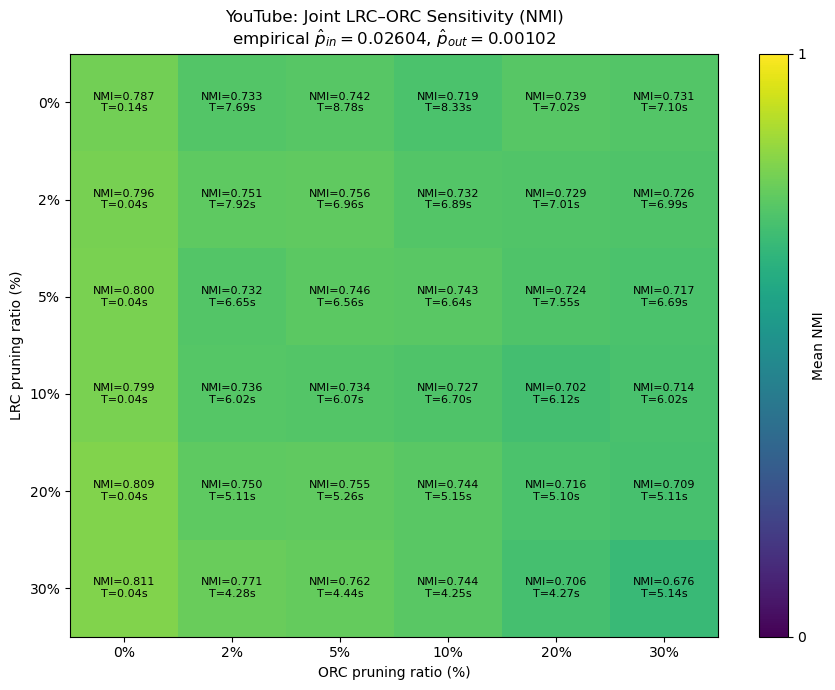}

    \textbf{(b) NMI}
\end{minipage}

\caption{Sensitivity of YouTube community-detection performance to
LRC and ORC-RF pruning parameters.}
\label{fig:appendix_youtube_heatmaps}
\end{figure}

\end{appendices}

\FloatBarrier

\section*{Code availability}

Code and reproducible analysis notebooks for the experiments in this study are available in the GitHub repository \cite{KolahdoozKolahdooz2026repo} at
\url{https://github.com/Avin-Kolahdooz/curvature-guided-community-detection}.

\section*{Acknowledgments}
This work was supported in part by the National Cancer Institute of the National Institutes of Health under Award Number U54CA272167. The content is solely the responsibility of the authors and does not necessarily represent the official views of the National Institutes of Health.

\section*{AI Usage Statement}
The authors used OpenAI's ChatGPT and Microsoft's Copilot MAI-Code-1.1-Flash for language editing, LaTeX/source-file editing, theoretical discussion, and refinement of Python code used in the computational experiments. All mathematical claims, code, numerical results, figures, references, interpretations, and final text were independently reviewed and verified by the authors, who take full responsibility for the manuscript.

\bibliographystyle{comnet}
\bibliography{sample}

\end{document}